\documentclass[manuscript,screen]{acmart}

\AtBeginDocument{%
  }

\setcopyright{acmlicensed}
\copyrightyear{2018}
\acmYear{2018}
\acmDOI{XXXXXXX.XXXXXXX}

\acmConference[Conference acronym 'XX]{Make sure to enter the correct
  conference title from your rights confirmation emai}{June 03--05,
  2018}{Woodstock, NY}
\acmISBN{978-1-4503-XXXX-X/18/06}

\usepackage{multirow}
\usepackage{booktabs}
\usepackage{threeparttable}
\usepackage{enumitem}
\usepackage[framemethod=tikz]{mdframed}

\usepackage{epigraph}
\usepackage{tikz}
\usepackage{tipa}
\usepackage{balance}
\usepackage{xtab}
\usepackage{mfirstuc}
\usepackage{todonotes}
\usepackage{soul}
\usepackage{xspace}
\usepackage{xcolor}

\usepackage{colortbl}
\usepackage[many]{tcolorbox}
\newtcolorbox{mybox}{
  breakable,
  colback=white,
  colbacktitle=white,
  coltitle=black,
  bottomrule=0pt,
  toprule=0pt,
  leftrule=3pt,
  rightrule=3pt,
  titlerule=0pt,
  arc=0pt,
}
\presetkeys%
    {todonotes}%
    {inline}{}
\usepackage{url}
\usepackage{relsize}

\definecolor{bgcolor}{rgb}{0.95,0.95,0.95} 

\newcommand{\qts}[3]{\emph{``#1''} (App: \textbf{#2} - Platform: \textbf{#3})  }
\newcommand{\qas}[3]{\emph{``#1''} (App: \textbf{#2} - Platform: \textbf{#3})  }
\newcommand{\qrs}[3]{\emph{``#1''} (App: \textbf{#2} - Platform: \textbf{#3})  }

\newboolean{showcomments}
\setboolean{showcomments}{true}
\ifthenelse{\boolean{showcomments}}
{ }

\ifthenelse{\boolean{showcomments}}
{ }

\ifthenelse{\boolean{showcomments}}
{ }

\newcommand{\noindentpara}[2]{\setlength{\parindent}{0pt}\paragraph{\emph{\textbf{#1}}}#2\par}

\begin{document}

\title{Unveiling Inclusiveness-Related User Feedback in Mobile Applications}

\author{Nowshin Nawar Arony}
\authornote{Both authors contributed equally to this research.}
\email{nowshinarony@outlook.com}
\affiliation{%
  \institution{University of Victoria}
  \city{Victoria}
  \state{BC}
  \country{Canada}
}


\author{Ze Shi Li}
\authornotemark[1]
\email{lize@uvic.ca}
\affiliation{%
  \institution{University of Victoria}
  \city{Victoria}
  \state{BC}
  \country{Canada}
}

\author{Daniela Damian}
\email{danielad@uvic.ca}
\affiliation{%
  \institution{University of Victoria}
  \city{Victoria}
  \state{BC}
  \country{Canada}}

\author{Bowen Xu}
\affiliation{%
  \institution{North Carolina State University}
  \city{Raleigh}
  \state{North Carolina}
  \country{USA}}
\email{bxu22@ncsu.edu}

\renewcommand{\shortauthors}{Arony et al.}

\begin{abstract}
In an era of rapidly expanding software usage, catering to the diverse needs of users from various backgrounds has become a critical challenge. 
Inclusiveness, representing a core human value, is frequently overlooked during software development, leading to user dissatisfaction. 
Users often engage in discourse on online platforms where they indicate their concerns. 
In this study, we leverage user feedback from three popular online sources Reddit, Google Play Store, and X, for 50 of the most popular apps in the world. 
Using a Socio-Technical Grounded Theory approach, we analyzed 22,000 posts across the three sources.
We organize our empirical results in a taxonomy for inclusiveness comprising 5 major categories: Algorithmic Bias, Technology, Demography, Accessibility, and Other Human Values. 
To explore automated support for identifying inclusiveness-related posts, we experimented with a large language model (GPT4o-mini) and found that it is capable of identifying inclusiveness-related user feedback.
We provide implications and recommendations that can help software practitioners to better identify inclusiveness issues to support a wider range of users. 
\end{abstract}


\keywords{Do, Not, Us, This, Code, Put, the, Correct, Terms, for,
  Your, Paper}

\received{20 February 2007}
\received[revised]{12 March 2009}
\received[accepted]{5 June 2009}

\maketitle

\section{Introduction}
As software usage continues to grow worldwide, an increasingly diverse user base is engaging with the applications.  
The diverse group includes individuals from various genders, regions, cultures, socio-economic backgrounds, political beliefs, people with physical and cognitive abilities, values, and educational backgrounds, among many others. 
However, software is often built for the ``average user'' \cite{savidis2006inclusive} and fails to adhere to the diverse user needs. 
For instance, X (previously known as Twitter), a widely used social networking app with over 390 million global users \cite{2023TwitterStat}, released an image cropping algorithm that automatically cropped images.
It focused on important parts, such as faces and text, to optimize space on the main feed and allow multiple pictures in a single tweet. 
However, users soon identified that the algorithm could only detect white faces and cropped out faces of black people \cite{2020faceCropTwitter}. 
The topic soon became trending as thousands of users joined the discussion.
Similarly, numerous other incidents have emerged from online user feedback \cite{li2022narratives},  highlighting the lack of inclusiveness in software. 

The feedback provided by users on online platforms (e.g. app reviews) has grown significantly in amount and has become important to software organizations. 
Software companies can identify areas of product improvement based on this feedback.
In this space, Crowd Requirement Engineering (CrowdRE) has become a popular area of study for identifying product relevant information from large volumes of user feedback on various online platforms such as app stores, social media, and forums. 
A growing body of research in ``end user human aspects'' has attempted to address and understand aspects such as gender and accessibility using CrowdRE sources such as App reviews \cite{shahin2023study, alshayban2020accessibility}.
Khalajzadeh \emph{et al.} \cite{khalajzadeh2022supporting} studied user feedback from Google Play Store and developer discussion from GitHub to understand human aspects related conversations from 12 open source apps.
The authors reported insights from discussions and concerns related to inclusiveness from both sources (31 from Google Play Store and 31 from Github).
Although insightful, open-source applications represent only a small portion of the many applications used in our society and, therefore, can result in limited representation of the diverse user needs.

Hence, there is a need for a more extensive exploration of concerns about inclusiveness from a larger, more diverse user base. 
With the increasing number of user feedback platforms (e.g., social media), diverse users may prefer to use different mediums due to different levels of engagement with particular online platforms \cite{tizard2022voice}. 
Thus, exploring a variety of sources of feedback can reveal more insights about inclusiveness. 
Finally, the growing amount of user feedback, while useful, represents a significant manual effort for software organizations, making the automation in identifying inclusiveness-related user concerns worth the effort.

Our study aims to fill this gap through analysis (using manual and large language models) of user feedback for 50 of the most popular mobile apps with millions of users, from Google Play Store, Reddit, and X.
Our work improves existing research by involving more data sources, different types of apps, and deeper qualitative analysis. 
We borrow from Khalajzadeh \emph{et al.}'s \cite{khalajzadeh2022supporting} definition of inclusiveness, namely \emph{issues related to the inclusion, exclusion or discrimination toward \textbf{specific groups of users}.} 
We use this definition and emphasize \textbf{groups of users} when conducting our qualitative analysis and identifying instances of users expressing exclusion from using an app or feature due to a lack of support for their needs.
Developing more insights regarding inclusiveness-related user feedback can assist software organizations in identifying critical issues that negatively impact diverse end users' ability to use an app whether it is due to age, location, or values.


Guided by the following research questions, we employed a Socio-technical grounded theory (STGT) approach \cite{hoda2021socio} to analyze the user feedback we collected from these multiple sources and software apps.

\begin{enumerate}[label=\textbf{RQ\arabic*},leftmargin=*]
    \item What are the different types of inclusiveness-related user feedback found on online sources? 
    \item How does inclusiveness-related user feedback differ across different sources of feedback?
    \item How effective are large language models in automatically identifying inclusiveness-related user feedback?
\end{enumerate}

We collected over 10 million posts and examined the inclusiveness-related user feedback both through qualitative analysis and by experimenting with large language models to automatically identify inclusiveness-related user feedback. Our study provides the following contributions:
\begin{itemize}[leftmargin=*]
    \item A two-layer taxonomy of inclusiveness based on user feedback from 50 of the most popular apps in a variety of types of software. The taxonomy comprises of categories of inclusiveness concerns such as \emph{algorithmic bias}, \emph{technology}, \emph{demography}, \emph{accessibility} and \emph{other human values}.
    
    \item Insights into the different inclusiveness concerns in different user feedback sources. 
    \item Insights and empirical results from using large language models to automatically identify inclusiveness-related user feedback from multiple sources, which companies can leverage to address the inclusiveness concerns of their end users.   
    \item A manually annotated dataset of inclusiveness-related user feedback that can facilitate future research and practice.
\end{itemize}
\section{Motivation}
Inclusive software is about designing with everyone in mind and considering the full range of human diversity \cite{2022DesigningMicrosoft}. 
Software systems are often built with a focus on functional or technical aspects as opposed to quality aspects \cite{werner2021continuously}, or requirements from diverse users \cite{khalajzadeh2022diverse}.
Consider a situation where a user of a specific software application cannot access content simply due to the user's geographic location. 
This lack of consideration of requirements in software systems from diverse users negatively impacts end users and excludes them from adequately using applications \cite{khalajzadeh2022diverse}. 
Our study answers the call for further research into ``inclusiveness'' user feedback from previous literature \cite{khalajzadeh2022supporting}. 
In this study we borrow the previous literature's \cite{khalajzadeh2022supporting} definition of inclusiveness: 

\epigraph{``This category (inclusiveness) covers the issues related to the inclusion, exclusion or discrimination toward specific groups of users.''}{}

However, previous literature focused their analysis on app reviews and issue comments \cite{khalajzadeh2022diverse, khalajzadeh2022supporting}, which missed out on other key sources of user feedback where end-users can voice their concerns regarding inclusiveness.
In addition to app reviews, Reddit posts or X posts offer users a space to express their opinions.
Reddit in particular offers end users an expanded commenting platform that is suitable for long-form discussions.

\begin{figure}[ht!]
  \centering
  \includegraphics[width=0.8\linewidth]{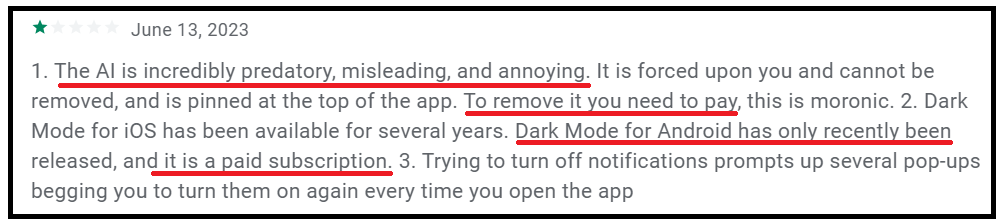}
  \caption{App review for Snapchat from Google Play Store. Underlined text indicates inclusiveness concern.}
  \label{fig:playstore}
\end{figure}

\begin{figure}[ht!]
  \centering
  \includegraphics[width=0.75\linewidth]{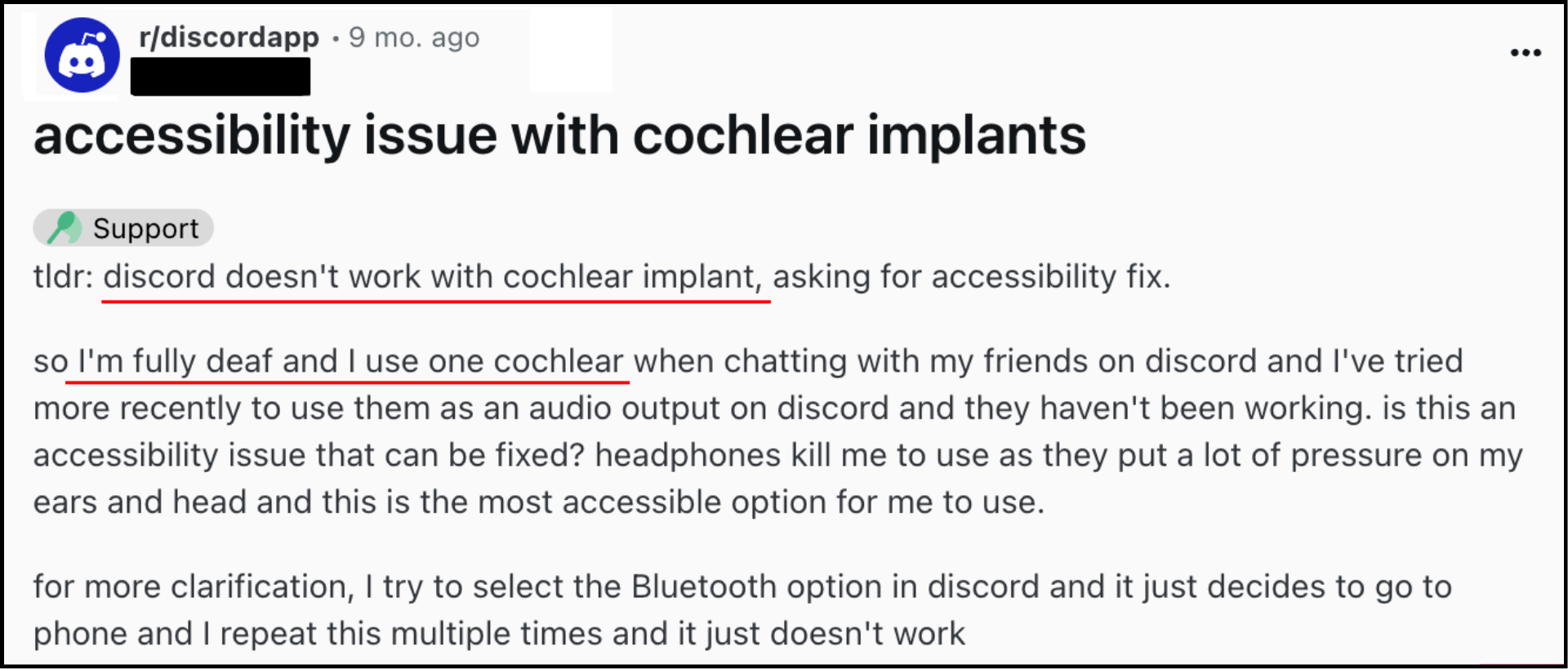}
  \caption{Reddit Post from Discord subreddit. Underlined text indicates inclusiveness concern.}
  \label{fig:reddit}
\end{figure}

\begin{figure}[ht!]
  \centering
  \includegraphics[width=0.75\linewidth]{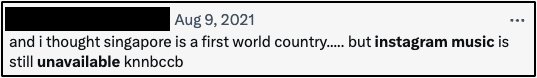}
  \caption{Post on X from a user.} 
  \label{fig:twitter}
\end{figure}

Figure \ref{fig:playstore} presents an example from Google Play Store where the user feels excluded from using dark mode on Android and removing the AI bot as they are not a paid subscriber, indicating technical and socio-economic restrictions.  
Similarly, Figure \ref{fig:reddit} shows an example from Reddit where a user describes their inability to chat with their friends on a popular communication app due to the lack of support for their cochlear implant. 
In Figure \ref{fig:twitter}, a user comments that they are excluded from accessing Instagram music due to being in Singapore, a geographic location not supported by the application. 
In these examples, users perceive that the groups they belong to such as users from different locations, accessibility needs, or users of certain devices are being excluded by the apps.


The examples also exemplify how diverse end-users have inclusiveness issues that companies should take into account to ensure a more inclusive user experience.
To further illustrate, consider an instance where an app does not support more than one language, in such case it would be excluding users who do not understand the provided language.
Using the aforementioned inclusiveness definition from previous literature, we embarked on an STGT study that explored the inclusiveness user feedback for mobile applications from the perspective of end-users.

\section{Related Work}
In this section, we describe the existing literature on inclusive software and user feedback.

\subsection{Inclusive Software}
The term inclusive software is related to the notion of ``universal access'' which implies software that is accessible and usable by everyone \cite{savidis2006inclusive}. 
The underlying philosophy behind designing an inclusive product is to ensure that the product can be used by as many different users as possible rather than excluding anyone \cite{pattison2006inclusive}. 
A study by Savidis and Stephanidis \cite{savidis2006inclusive} highlights the significance of providing the necessary tools to support inclusive software design and development.
They indicate that an important aspect of inclusive software development is identifying user requirements that emerge from interaction with the software. 

Conventionally, software is developed with a focus on the average user, and requirements for the software are developed with this perspective. Recent studies, however, highlight the need for more inclusive software.
For example, although software is primarily intended to be neutral, software interfaces often contain stereotypical visual components that negatively impact many users' sense of belonging \cite{metaxa2018gender}.  
Burnett \emph{et al.} \cite{burnett2016gendermag} revealed that problem-solving software is developed with a perception that users will adopt the features through tinkering.
However, statistically, these features are preferred by men than by women, making the software less inclusive for women.

There are some prior works toward building more inclusive software, particularly focusing on gender inclusion.
Nunes \emph{et al.} \cite{nunes2021conceptual} proposed a conceptual model for gender-inclusive requirements that involves creating a gender-inclusive requirements document. 
The document supports practitioners in integrating the model into the requirement elicitation process.  
Upon evaluation of the model, they found 83.9\% positive response in terms of the usefulness of the model \cite{nunes2023gire}. 
The GenderMag (Gender Inclusiveness Magnifier) method developed by Burnett \emph{et al.} \cite{burnett2016gendermag} uses personas encapsulating five facets of gender differences to analyze gender inclusivity in software.
An empirical investigation of GenderMag identified biases in an industrial software and helped derive design changes that improved the inclusiveness of the software \cite{vorvoreanu2019gender}.
Guizani \emph{et al.}, in their study, proposed a Why/Where/Fix approach to find and fix inclusivity bugs in an Open Source Software project.
They reported their approach reduced inclusivity bugs by 90\%. 

While these studies focus on addressing gender-related concerns, the concept of inclusion extends beyond gender.
Recent studies have indicated that to make software more inclusive, software companies need to better understand human aspects such as age, emotions, personality, human values, gender, ethnicity, and culture \cite{grundy2021impact, grundy2021addressing}.
There are various ways to understand the different human aspects of diverse users.
For example, co-design or participatory design techniques where users are invited to participate and provide feedback during the design process \cite{donetto2015experience}. 
However, as software grows and becomes more prevalent around the world, it becomes difficult for companies to conduct such design sessions. 
In such cases, CrowdRE (Crowd Requirement Engineering) \cite{groen2015towards} techniques can be leveraged.
In our study, we complement current research and use the CrowdRE method of exploring inclusion from an end-user perspective in comparison to prior studies that focused on understanding inclusion from a developer or designer perspective.

\subsection{User Feedback}
Prior literature has shown that user feedback is beneficial for continuous improvement of software quality \cite{pagano2013user}. 
User feedback from the online platforms, i.e., from the ``Crowd'' \cite{groen2017crowd},  has been studied to identify a variety of user needs.
One common type of user feedback found in app reviews \cite{maalej2015bug}, Twitter (now referred to as X) \cite{williams2017mining}, and Reddit \cite{iqbal2021mining} are feature requests and bug reports.

More recently, Fazzi \emph{et al.} \cite{fazzini2022characterizing} analyzed 2,611 app reviews from 57 COVID-19 apps and found nine categories of human aspect related discussions that impact software usage.
The authors implied that these human aspects are not always taken into consideration and should be addressed during development.  
Another study on 1,500 top free Android apps more focused on accessibility issues revealed that the majority of these apps contain significant problems that prevent individuals with disabilities from using the apps \cite{alshayban2020accessibility}. 
The study demonstrated that various sub-categories of human aspects are identifiable from user feedback, which can raise awareness amongst developers and companies, enabling them to incorporate these insights during development. 
Likewise, Arias \emph{et al.} \cite{reyes2022accessibility} demonstrated that users often express accessibility concerns in non-technical terms in mobile app reviews, which can be mapped to accessibility guidelines, such as WCAG 2.1.

Shahin  \emph{et al.} \cite{shahin2023study} conducted an analysis of gender related discussions on app reviews and found six major categories: AppFeatures, Appearance, Content, Company Policy and Censorship, Advertisement, and Community. 
In addition, they automated the identification of gender and non-gender related discussions and acquired an F1-score of 90.77\%. 
Li \emph{et al.} \cite{li2022narratives} obtained 4.5 million posts from Reddit and found 9 significant topics related to privacy concerns. 
Similarly, Olson \emph{et al.} \cite{olson2023along} examined 586 subreddit communities and identified discussions on ethical concerns from end users regarding social platforms. 
Tushev \emph{et al.} \cite{tushev2020digital} analyzed user feedback from sharing economy platforms Uber, Airbnb, and TaskRabbit to identify and classify digital discrimination concerns. 
Obie \emph{et al.} \cite{obie2021first} analyzed 22,119 app reviews from the Google Play Store using natural language processing techniques, identifying that 26.5\% of the reviews indicated user perceived violations of human values where benevolence and self-direction were the most frequently violated categories. 

Khalajzadeh \emph{et al.} \cite{khalajzadeh2022supporting} manually examined 1,200 app reviews and 1,200 GitHub issue comments for 12 open source projects and characterised human aspect issues, finding three major categories: App Usage, Inclusiveness, and User Reaction.
While insightful, they represent only a starting point with respect to inclusiveness issues as the open source apps represent a small number of mobile app users.
Hence, we still lacked empirical insights about inclusiveness in mobile apps from a broader context of apps and users. 

To address the gap, we conduct an analysis of inclusiveness-related user feedback from 50 of the most popular software applications in the world, using three popular sources such as Reddit, Google Play Store, and X
We do not consider GitHub as a source in this study because it primarily represents the developer's perspective rather than that of the end users.
Lastly, in contrast to prior work employing an iterative, open coding analysis method, we employ a socio-technical grounded theory \cite{hoda2021socio} approach in our study.

\section{Methodology}
In our study we collected user feedback from three popular platforms and employ Socio-Technical Grounded Theory (STGT) \cite{hoda2021socio} method.
We chose STGT to guide our methodology because 

\epigraph{``STGT is an iterative and incremental research method for conducting socio-technical research using traditional and modern research techniques to generate novel, useful, parsimonious, and modifiable theories.'' \cite{hoda2021socio}}{}

Unlike classic GT (Grounded Theory ) or Straussian GT, Hoda stressed that STGT facilitates software engineering research using modern research data, tools, and techniques such as mining publicly available software code repositories \cite{hoda2021socio}. 
Our research methodology adheres to STGT data analysis steps including open coding, constant comparison, and basic memoing, which resulted in a taxonomy of inclusiveness-related user feedback.

\subsection{Data Collection} 
We collected data from three popular online sources of user feedback: Reddit, Google Play Store, and X. 
We chose the three sources as prior studies have successfully found software relevant information, such as bugs and features from these channels \cite{maalej2015bug, williams2017mining, iqbal2021mining}.
Reddit is popular for having a high character limit that allows users to engage in elaborate discussions.
A single Reddit post has the room to provide extensive details about a particular issue that otherwise is not available in comparable feedback sources.
Google Play Store offers app users to leave reviews about any app, which is useful for software organizations to elicit concerns regarding any particular app.
In contrast, X is well known as one of the most popular social media platforms, which supports short form textual user discourse about any particular topic.
X has been shown to provide requirement relevant information for organizations to analyze \cite{guzman2017little}.

We collected a list of 50 of the most popular apps from Google Play Store.
These apps have come from various domains and are actively installed by a diverse group of users from across the world.
This list is used to scrape the data for Reddit and X as well, thereby giving us a unified range of apps.
For Reddit, we use a publicly available dataset \cite{redditData} and obtain over 380,000 Reddit posts.  
Next, we collect 9 million app reviews from Google Play using the library Google Play Scraper.\footnote{\url{https://github.com/JoMingyu/google-play-scraper}} 
Lastly, we use the snscrape library\footnote{\url{https://github.com/JustAnotherArchivist/snscrape}} to scrape over 800,000 discussions from X. 

Then we filtered the original data by removing empty posts.
and filtering out any post that has less than three words as posts that cannot satisfy this criterion most likely do not provide meaningful information. 
We were left with over 3.7 million app reviews, 824 thousand X posts, and 359 thousand Reddit posts.

\subsection{Qualitative Analysis}\label{sec: method-analysis-method}
To analyze the data, we followed the basic data analysis technique from Socio-Technical Grounded theory (STGT) \cite{hoda2021socio}, and which allows us to establish important categories or initial hypotheses from the user feedback.
Our goal was to develop an understanding of inclusiveness-related to user feedback from the end user perspective. 
The STGT basic analysis step consist of open coding, constant comparison, and basic memoing.
For the open coding, we randomly sampled feedback from the three sources for all 50 apps. 
We used random sampling to minimize selection bias \cite{infante2018reflection} and ensure that the labeled data adequately represented the diversity of inclusiveness-related feedback in the dataset, providing a starting point for the analysis.
To guide our open coding process, in line with STGT, we conducted a literature review to identify the existing understanding of inclusiveness in software engineering (as outlined in the Related Work section), and used the following definition of inclusiveness \cite{khalajzadeh2022supporting}. 

\epigraph{``This category (inclusiveness) covers the issues related to the inclusion, exclusion or discrimination toward \textbf{specific groups} of users. It includes issues related to the age, gender, and socioeconomic status of the users.''}{}

As per the definition, we labeled any user feedback as inclusiveness when we found a user describing that they were unable to use an app or its features, and perceived that this was happening because they belonged to a specific group such as being a person with a disability, being from a specific location, or even having particular devices.
Two members of the research team analyzed the randomly sampled data and assigned an inclusiveness or misc (non-inclusiveness) label to each post. 
When we label a post as inclusiveness, we then included a code based on the characteristics of the feedback.
We assigned one code to each inclusiveness feedback because our goal was to identify the predominant inclusiveness concern in each post and understand the most pressing issues indicated by users.
Moreover, labeling with a single code simplified the data analysis process and allowed for clearer interpretation of results without the complexity that multi label coding can introduce \cite{saldana2021coding}. 
For example, the inclusiveness quote \emph{``Why ain’t Reels available for everyone in every country and Instagram music? It sucks not being able to hear the sounds some [people] post just because IG music isn’t available in my region.''} was coded as \textit{location} as the user expressed frustration about feeling excluded due to their geographic location.

\begin{table*}[h!]
    \centering
      
    \caption{Example Open Coding of Raw Quotes and Mapping to Codes and Categories}
    \small
    \begin{tabular}{p{10cm}p{1.5cm}p{1.5cm}} \toprule
       Raw Quote indicating an \textit{inclusiveness}-related concern& Code (sub category) & Category\\ \midrule
         Why ain’t Reels available for everyone in every country and Instagram music? It sucks not being able to hear the sounds some [people] post just because IG music isn’t available in \textbf{my region}. & Location  &  Demography  \\  \midrule
        Feedback has been given to @Zoom about this innumerable times. At this point, the only thing I can take from this is that they don’t care about how damaging their shitty captions are to \textbf{disabled and neurodivergent
people}. All @Zoom does is enable microagressions. & Visual  &  Accessibility  \\  \midrule
        So I looked in my google play store and
the app is no longer compatible with \textbf{my device}??? LOL Been having issues left and right with the app and an update was
denied lol I have a \textbf{Pixel 7}....how is this no longer compatible? & Device & Technology \\
         \bottomrule
    \end{tabular}
    \label{tab:coding}
\end{table*}

Any time a new code emerged for the inclusiveness user concerns, the two human annotators met and discussed the implications of the new code. We used constant comparison method to compare the derived codes across all three user feedback sources and to uncover the underlying patterns and relations between the codes. Our analysis yielded 14 codes under inclusiveness, and which we further grouped into 5 categories.
Table \ref{tab:coding} illustrates examples of the code and categories. 
We employed the basic memoing technique to document the reflections on the emerging codes and categories. A significant part of STGT memoing, encapsulates the researchers' thoughts, enabling a systematic development of categories from the initial codes. 

Another important aspect of STGT is theoretical saturation, which means ``when data collection does not generate new or significantly add to existing concepts, categories or insights, the study has reached theoretical saturation'' \cite{hoda2021socio}. 
Therefore, the two human annotators continued the labeling process until no new categories emerged and code definitions became stable (i.e.saturation). 
For Reddit and Google Play Store, we observed no additional insights emerging from the last 200 posts. For X, due to the presence of irrelevant discussions in the data, saturation was reached after we coded 500 additional posts. 
In total, we labeled 4,420 Reddit posts, 4,962 from Google Play Store, and 13,500 from X.

 
We present categories from our analysis in the form of a two-layer \emph{taxonomy of inclusiveness}, with five categories forming the primary layer and the 14 codes distributed within each category as a sub-category. 
The labeled data and memoing can be found in the replication package \cite{replication_package}.

\subsection{Automated Analysis} \label{sec: classification-method}
To analyze the effectiveness of automatically identifying inclusiveness-related user feedback, we experimented with GPT4o mini, which is one of the state of the art large language models. 
Using GPT4o mini,\footnote{https://openai.com/index/gpt-4o-mini-advancing-cost-efficient-intelligence/} we conducted binary classification: \emph{inclusiveness} and \emph{non-inclusiveness}.
We selected GPT-4o Mini due to its balance between computational efficiency, cost, and performance. 
GPT-4o Mini is suitable for diverse data types and its cost-effectiveness and reduced computational requirements facilitate experimentation.
Other studies have already explored leveraging GPT3.5 and GPT4 for similar text classification activities. 
For example, papers have investigated ChatGPT for stance detection of social media \cite{zhang2023investigating} and financial text classification \cite{li2023chatgpt}. 
Recall in Section \ref{sec: method-analysis-method} we collated a human annotated set of user feedback.
This labeled set served as a ground truth for us to train and evaluate the model.

To assess the effectiveness of GPT4o mini, we implemented three different approaches. 
\begin{itemize}
    \item Zero shot Learning: This approach evaluates the model’s ability to classify inclusiveness-related content without prior exposure to context specific examples. 
    \item Few shot Learning: This approach is provided with 5 labeled examples, one from each category, to guide its classification process. This approach assesses how minimal contextual information can improve the model’s understanding to identify inclusiveness issues.
    \item Fine tuning: This approach involves training GPT-4o Mini on a substantial subset of labeled data specific to our inclusiveness taxonomy. Fine-tuning aims to adapt the model’s parameters to our dataset, thereby improving its performance in detecting inclusiveness-related feedback.
\end{itemize}

We evaluated the three approaches on the same test dataset containing 1000 user feedback. The test dataset is imbalanced (5\% inclusiveness, 95\% non-inclusiveness), reflecting the imbalanced nature of inclusiveness-related user feedback.
For Zero-Shot learning and Few-Shot learning, we directly evaluated them with the test dataset. 
For Fine-Tuning, we first trained on a balanced dataset of 1,200 user feedback (50\% inclusiveness, 50\% non-inclusiveness) before evaluating on the test dataset. 
We prepared a balanced training dataset, as an imbalanced dataset can cause machine learning models to prioritize the major class and bias against the minor classes \cite{akbani2004applying}.
Due to space considerations, we did not include our GPT4o-mini prompts here in the paper, however, they are provided in the replication package \cite{replication_package}.
We report the performance of the classifiers and use 3 widely used evaluation metrics: precision, recall, and F1-score.
We compute the macro average score for all three metrics 
because the macro calculates scores independently for each class and then takes the average across all classes. 
Macro average treats all classes equally, regardless of their size. This makes it particularly useful for evaluating the model’s performance on minority classes, as it does not favour the dominant or majority class.

In the sections that follow, we present the empirical results of our study.

\section{RQ1: A Taxonomy of Inclusiveness} \label{sec:taxonomy}
\begin{figure*}[ht]
  \centering
  \includegraphics[width=0.95\linewidth]{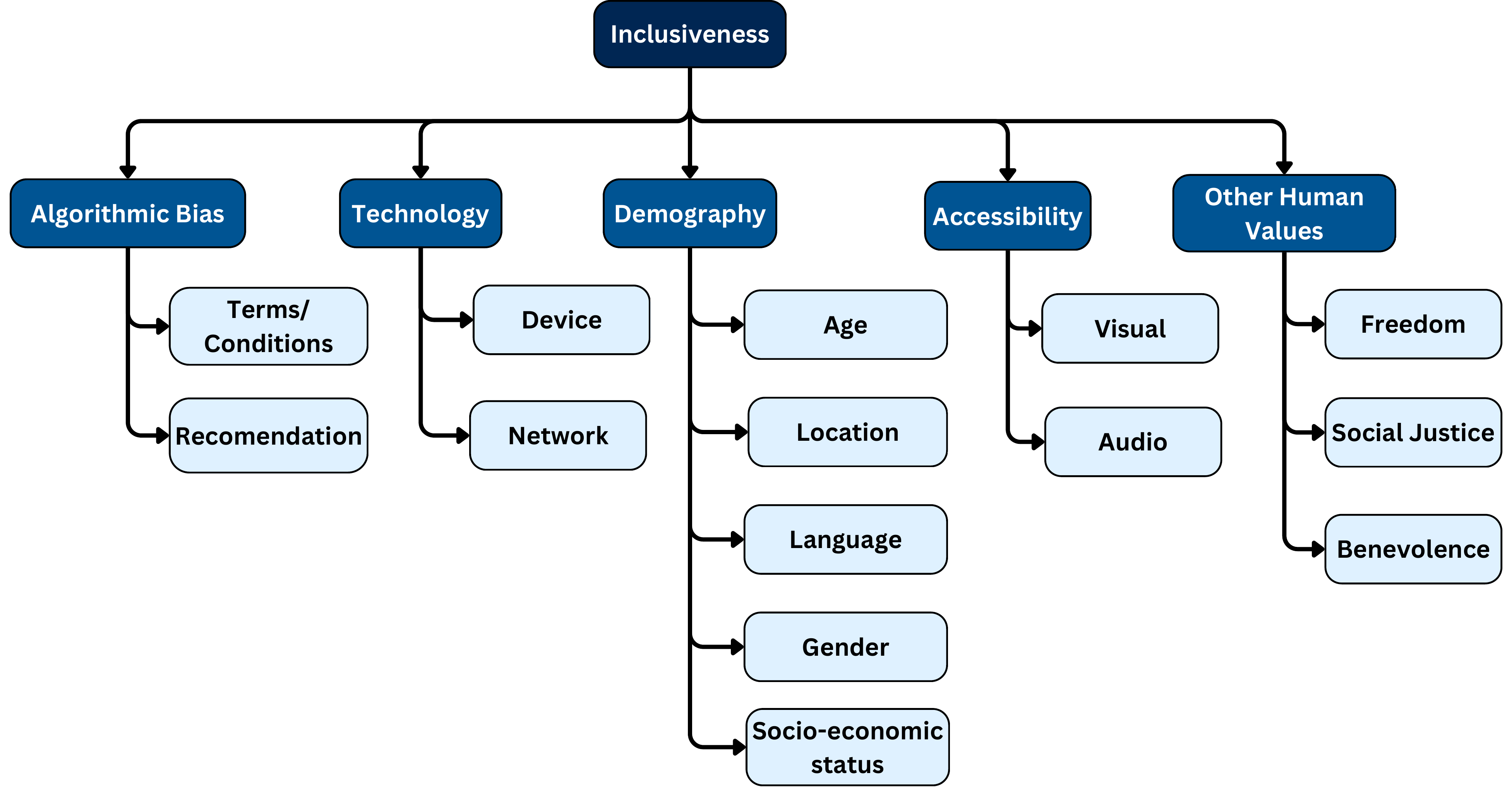}
  \caption{Taxonomy for inclusiveness-related user feedback from an analysis of Reddit, Google Play and X }
  \label{fig:taxonomy}
\end{figure*}

To answer our first research question \textit{What are the different types of inclusiveness-related user feedback found on online sources?}, our in-depth analysis of 4,647 discussions from Reddit, 4,949 from App reviews, and 13,511 from Twitter identified a total of 1,211 inclusiveness-related posts: 712 from Reddit, 377 from Google Play Store, and 116 from X (Twitter). 
Using STGT we derived 5 categories and associated sub-categories of inclusiveness from the three sources. We integrated them into a \emph{taxonomy of inclusiveness} from end-user feedback, as illustrated in Figure \ref{fig:taxonomy}. 
In this section, we describe each category with examples from our analysis, indicating in brackets the percentage from each source (while R refers to Reddit and A to app reviews from Google Play Store, X symbolizes results from X (previously known as Twitter)). 




\subsection{Algorithmic Bias}  
Algorithmic bias includes any user feedback in which a user describes unfair treatment towards them or a group they belong to during the app usage due to algorithms in a software application.
It is one of the most common categories of all-inclusiveness concerns (R: 21.3\%, A: 26.7\%, X: 13.5\% ). 
Algorithmic bias ``occurs when the outputs of an algorithm benefit or disadvantage certain individuals or groups more than others without a justified reason for such unequal impacts'' \cite{kordzadeh2022algorithmic}. 
As algorithms play a growing role in software systems from content recommendation to enforcement of terms and conditions, biases in these algorithms may cause inequality and have  a significant impact on user experience.
We found that algorithmic bias occurs in two sub-categories: \emph{\textbf{Terms/Conditions}} and \emph{\textbf{Recommendation}}.

{\noindentpara{Terms/Conditions}{We found users frequently complaining about the inconsistent enforcement of terms and conditions. 
This bias is often observed in content moderation practices, where automated decisions determine what content is allowed, flagged, or removed. 
Such bias is also commonly observed in account restrictions, where AI systems decide if an account violated app rules and warrants consequences. 
Studies have shown that biases in enforcement can lead to perceived unfair treatment, eroding trust in an application 
\cite{pan2022comparing, godfrey_new_2024}.
}}

One Reddit user expressed their frustration over the constant hateful comments about their unusual last name from others on Facebook. Despite reporting the hate speech, the platform did not find the comments to violate the community standards. 
\qrs{I have an \textbf{unusual last name}, people on groups will constantly make some hateful remark out of the blue. I report it and Facebook community standards never find anything wrong with their comments.
I have been called all kinds of names on FB and nothing ever gets taken down. Yet one time I responded to someone and got a warning for making a remark and defending myself. Anyone else deal with this or have issues with their community standards?}{Facebook}{Reddit}
Additionally, the Reddit user who experienced hate speech was dealt with a warning for defending themselves. This experience led the user to feel that the platform is biased and lacks inclusiveness toward specific user groups like theirs.

Similarly, the feedback from X indicates that the platform shadow bans pro-Black Americans only.
\qts{Twitter is aggressively making an effort to shadow-ban \textbf{pro-black accounts}. Once my tweet regarding [name] received close to 100,000 likes and 20 million impressions, they began actively targeting me and shadow-banning me. The status quo is endangered by our voice.}{X}{X}

Likewise, freelancers on Fiverr complained that they received unfair restrictions of their accounts without any warning or justification. 
\qas{Such a racist and corrupted application Don't use this platform. They are blocking so many hardworking \textbf{freelancers} without warning. Just boycott them. If you got blocked your account, just stop using their service and tell others not to use these corrupted Fiverr apps}{Fiverr}{Play Store}
In this case, a freelancer not only felt that their group was frequently targeted, but also called for others to boycott the app.

All of these quotes exemplify issues with how certain groups of end users feel a lack of inclusiveness toward using the apps fairly and equitably. 
In particular, there is little transparency regarding how the content moderation and account restriction systems are designed to enforce terms and conditions. 
Biases in these systems can result in unjust bans. 
For example, the feedback from X suggests that users from minority groups are unfairly targeted for shadow bans.
Previous works suggest that biases in training data used to train AI algorithms that make automated decisions may contribute to this discrepancy, leading to different impacts on different user demographics \cite{callahan_algorithms_2023}.

{\noindentpara{Recommendation}{Similar to terms and conditions, users expressed frustration with the automated recommendation systems in apps.
Recommendation systems are among the most important and ostensible features in apps because they power the content suggestions in movies, news, or any social media apps based on past user behavior and demographic information. 
However, we found users concerned about biases that exist in these systems leading to negative experiences and feelings of exclusion from using apps. 
}}

Users start complaining when they feel that these machine learning systems fail to consider their preferences, which occurs when an app unexpectedly and continuously suggests content that the user does not want to see. 
When this happens and is not rectified by the app developer, it may cause the user to feel that the app is intentionally mistreating the user.
As one user on Reddit pointed out, they feel that Facebook inaccurately labeled their race and as a result constantly bombards them with advertisements that are geared towards other specific ethnicity. 
\qrs{As the title says, I am \textbf{White}. However, Facebook 100\% think \textbf{I'm Black or Native American}. It keeps showing me ads which are specifically targeted towards African-Americans or Native Americans. I clicked ``never show this page again" on all I see, but still some pop up now and then. I don't know why it does this. I don't know any Black or Native American people on Facebook and have no interest in anything the ads are showing. How can I get Facebook to stop thinking I'm Black?}{Facebook}{Reddit}
Despite indicating to the Facebook advertisements to not show them again, the user still sees the same type of content.

Similarly, a user on X expressed significant frustration at receiving content recommendations on YouTube that are hateful. 
\qts{Tell my why i keep getting \textbf{queerphobic} or \textbf{racist} youtube shorts recommended? they don’t have anything to do with what i watch and it just pissed me off so bad}{Youtube}{X}
In this case, the user indicated that they do not consume content that relates to the hateful shorts and is confused as to why this is the content recommended to them. 

Additionally, users expressed concern about the recommendation systems in video-on-demand streaming services, worried that the recommendations are biased. 
\qas{Whoever runs the categories section is RACIST. You have black stories. You have Latin and Hispanic stories. Okay, races, , where are the \textbf{Asian stories}? Where are the \textbf{white stories}? What makes those 2 categories so important You have to include them as separate? Are all stories just human stories. Is racist [explicit] promoting segregation}{Hulu}{Play Store}

These quotes exemplify the issues that manifest from recommendation systems that result in biased content suggestions. 
When AI algorithms make wrong assumptions about a user's demographics such as ethnicity, the content does not align with user interests and potentially causes the user to feel alienated from the app, creating a feeling of lack of inclusiveness.

\subsection{Technology}
This user feedback refers to user concerns about experiencing exclusion from an application due to restrictions enforced by the developer, i.e. exclusion from a software or feature, due to certain technology restrictions enforced by the developer.
One other most prevalent category out of the 5 (R: 25.9\%, A: 26.3\%, X: 26.6\%).
These limitations arise when developers insist on users using a particular \emph{\textbf{device}}, or \emph{\textbf{network}}. 
Inadvertently, developers may exclude the group of users by specifying certain devices or placing certain demands on network speeds.

{\noindentpara{Device}{Users are often excluded from using an app when features for the app no longer work on an older device. This is particularly troubling for users who may not have the ability to switch to a newer device or struggle to learn to use newer devices. }}

For example, one user explained how their elderly parent has an older computer and would like to use the sharing features, however, such features are not supported on her device.  
\qts{What do you think of my soon to be \textbf{91 yr old mom}? She’s got an old like 15 yr old Mac that she does Facebook and apps for some of her senior issues. Yet she’d like to zoom some library and civic discussions her old Mac doesn’t support. She lives on her own}{Zoom}{X}
Elderly folks who are in the same situation and use older devices will have the same problem of not being able to use the other sharing features.   
Likewise, a user complained on the Play Store that they cannot use the background feature on the video conferencing application as they are on an older Android device.
\qas{Can't use background at an \textbf{old android,} prefer use gmeet than this.}{Zoom}{Play Store}

Another user complained that their relatively new phone is no longer compatible with a popular crypto exchange application to the point where they can no longer even update the application.
\qrs{So I looked in my google play store and the app is no longer compatible with \textbf{my device???} LOL
Been having issues left and right with the app and an update was denied lol
I have a \textbf{Pixel 7}....how is this no longer compatible?}{Coinbase}{Reddit}

In all these examples, the user group experiencing a lack of inclusiveness includes users with older or unsupported devices, as well as seniors. 
From the examples, we see that the lack of support for apps or features in various devices is a recurring problem echoed by users across Reddit, X, and app reviews.
The impact on an end user of a device not supported by an app is quite significant. 
It often means at minimum the user cannot use a feature, or worse, the user cannot use the entire app.

{\noindentpara{Network}{In addition to devices, we observed instances of users criticizing certain apps not fully functioning on their networks, whereby users are excluded from using these apps.
These user groups include people with specific internet setups or bandwidth limitations, possibly due to their internet quality.
}}

One app store user recounted how a popular video streaming application can never work on their wifi. 
The application cannot connect to the network despite the user having fast internet. 
\qas{the app \textbf{never works with my wifi}. I have to Initiate the app via cellular data, load the show I want to watch, and THEN initiate my wifi connection. I do not have slow internet.}{Hulu}{Play Store}

Likewise, one user on Reddit complained about how one messaging app is the only app that refuses to connect on their Wifi.
\qrs{Me and my girlfriend have been having this issue for a week or so. We both have iPhone 14 Pros and when \textbf{we are connecting to our WiFi, WhatsApp can't connect}, as soon as we switch to data it works absolutely fine.
No either device or App is having any issues with our wifi/internet connection it seems to only exist with WhatsApp, I have tried deleting the app/reinstalling it, I have tried forgetting the wifi network on my phone and rejoining it, and I have tried hard restarting my phone but nothing seems to be fixing the issue, any help appreciated.}{Whatsapp}{Reddit}

In both of these instances, users are forced to use the apps on their mobile networks, which can significantly drain their wireless data plan. This can be an issue for users who reside in places that have higher internet costs or low wireless data speeds.

\subsection{Demography}
One of the most common categories of inclusiveness-related user feedback pertains to user demographics. 
In particular, users expressed concerns related to various demographic factors including \textbf{\emph{age}}, \textbf{\emph{gender}}, \textbf{\emph{language}}, \textbf{\emph{location}}, and \textbf{\emph{socio-economic status}} (R: 19.2\%, A: 19.2\%, X: 18\%). 
Early research on human-centric issues in open source software found issues related to location, language, and socio-economic status \cite{khalajzadeh2022supporting}, but our findings expand on this with the addition of age and gender. 
Failure of app developers to consider one or more of these demography factors can exclude users from using apps to the full extent of their features. 

{\noindentpara{Age}{Different age groups have varied motor and non-motor abilities, which impacts how they interact with different apps and limits their ability to adapt quickly to user interface changes \cite{yu2020maps}. Moreover, users may be excluded from using an app based on age-related policy violations.}}

For example, one user on Reddit criticized the confusing and ambiguous steps needed to enable original sound on Zoom.
The user explained that the steps are repetitive, tricky, and difficult even for tech-savvy users.
\qrs{I do music lessons with [Zoom] and it is such a buzz kill every time I get a new student to have to train them to first find the setting to ``enable the option to turn on original sound" // and then find the SECOND hidden menu where you can actually ``turn on original sound."
It's a different process on iPhones than on iPads. it's different on androids. it's different on MacOS. It's different in windows. It's different in the web browser. It's different if youre device is in vertical vs horizontal orientation.
This is a such a fundamental feature but the process to enable it is so god damn unintuitive and cumbersome procedure that techie 30 year olds often struggle with it. Let alone the \textbf{young children}, and \textbf{elderly} clients that make up a bulk of my business.}{Zoom}{Reddit}
Users who fall outside the tech-savvy demography struggle to effectively toggle the correct audio setting. 

Similarly, one user on X complained about the difficulty that old people have using the Shopify website due to the font size. 
\qts{Shopify tip: Increase your websites font size. Most \textbf{old people} have terrible vision and can’t see sh*t.}{Shopify}{X}

There was another user who expressed frustration over age verification that limited their ability to use a short video social media application.
\qas{I had submitted a claim to get my age verification done after not being able to have the option to add my birthday even though I did it when signed up. Reported it. Team followed up. I then did as they asked. I submitted a second claim and got no response. They ghosted me. It's absolutely ridiculous I followed instructions only to never hear back. Can't view age restricted videos but \textbf{im over 18.} I have now told everyone to avoid this app.}{Tiktok}{Play Store}
The user critique is that they were unjustifiably age limited, which prevents full access to videos on the platform.

{\noindentpara{Location}{Another common sub-category is location, which encapsulates any user feedback about exclusion from using an app based on a user's location.
Location may refer to a user's country, city, province, state, or any other aspect to do with the geographic coordinates. }}

For example, one user expressed outrage that an important feature from a popular social media application was not available simply due to the user's region.
\qas{Why ain't Reels available for everyone in every country and Instagram music? It sucks not being able to hear the sounds some [people] post just because IG music isn't available in \textbf{my region.}}{Instagram}{Play Store}

Similarly, one user on Reddit questioned how they can conduct phone verification when their country is not even listed as an option to receive phone verification.
\qrs{My account got locked and it asked me to verify my phone and email but \textbf{my country code} isn't even an option how am I supposed to verify? Country code is South Africa (+27) }{X}{Reddit}
\qrs{In \textbf{new zealand} you cant withdraw any amount of money. Any amount of money i try to withdraw (tried 20\$) will say "does not meet the minimum amount required to withdraw". I now have money sitting there forever, and need to do some other sales but cant if i cant withdraw the money.
On top of this, paypal (that ive been able to see) only does support via phone, on numbers that you cannot call from \textbf{new zealand.} So i essentially have no contact for support, and hence why im asking for help here.}{Paypal}{Reddit}

In all these examples, users complained about a variety of issues related to exclusion from using the apps due to restrictions based on their location.
For Instagram, the user feels excluded from the app as music is not available in every country, and the user's country is not on the list of exempt countries. 
Similarly, a user of Paypal in New Zealand laments that they cannot withdraw money due to location restrictions. 
In addition to this, Paypal also cannot request support either as Paypal does not provide a local phone number for people in New Zealand.

We observe similar location problems when it comes to account opening or closing. 
An Etsy user complains that they cannot open an Etsy shop because they cannot use codes from the authenticator app, but this was their last opportunity after exhausting all other available attempts to open the account. 
\qrs{Hello! I am trying to open an Etsy shop, but when I try to set up my shop security, by the authenticator app, I just can't... Every code I entered was a correct one, but Etsy said they weren't valid. Now I have too many failed attempts and do not know how to proceed... The other methods do not work for \textbf{my country}... Any tips? Has anybody dealt with this before? Thank you...}{Etsy}{Reddit}

We also see location issues that occur from sudden shifts in geographic restrictions. 
A feature that used to work for users is modified and no longer works for users from a country or state. 
\qrs{Trying to buy with N26 card from France Used to buy without any problem, now I get this:
``We currently do not support the bank cards issued in your country, but we are working on making this a possibility as soon as we can."  
Did they ban Germany from Binance? What's happening?}{Binance}{Reddit}

{\noindentpara{Language}{Similar to location, we find an assortment of user feedback about lack of inclusiveness-related to user language, where a user's preferred language is not supported by an app.}}

One user criticized a video streaming application for the lack of language support for their region. 
\qts{Dear @PrimeVideo, can you explain to me why, for the same show, same episode, I have on my phone only one other language (not even \textbf{my location's language}) available somehow and NO subtitles available in whatever language, and on the Windows app I have like 40 choices for both?}{Amazon Prime Video}{X}
In this case, users in this region are collectively excluded from being able to watch videos with the subtitles of their home language. 

Alongside this example is another instance from Reddit where a user complained about their disappointment towards another video streaming application regarding the lack of support for additional languages and subtitles. 
\qrs{I recently got back on Hulu after nearly three years and figured after all this time they would have finally added \textbf{multiple languages} to their shows like Netflix and HBO Max, but nope. Still only select shows with subtitles in Spanish. What gives? Even Disney+ has multiple languages available for nearly all their content.
Same with their live TV not offering SAP. I feel like after all this time with all this competition, someone would have brought up adding multiple languages to their shows.}{Hulu}{Reddit}

The language issue is not unique to video stream applications as we also observed similar issues in social media applications. 
\qas{One star because instagram doesn't have Albanian language}{Instagram}{Play Store}
Automated language recognition software may not achieve 100\% accuracy and users are annoyed when posts from the wrong language is suggested. 
\qas{Usually good, but the change in notification is annoying and confusing, a lot of creators are harassed and receive strikes when they are being harassed, the language configuration is not able to detect the language in content and often send me content in languages I signaled as not interested. ...}{TikTok}{Play Store}

We observe from these examples that criticism is directed at the choice of languages in the app and that their choice of content is excluded from the app. 
End users cannot use an app if their preferred language of choice is not supported by the app developers, leading to users feeling excluded from the app. 

{\noindentpara{Gender}{In addition to the aforementioned demography factors, we also found issues related to gender inclusiveness.}} 
One user complained on X about news posting about violence facing one gender resulted in account suspension. This led to the user speculating whether this was a result of intentional suppression of this type of news.
\qts{Suspended for pointing out the crimes of men committing violence against women again. @Twitter allows for those crimes to be posted but \textbf{suspends/bans women} who point them out. If we can’t speak, then no one knows \& twitter wants to make sure no one knows.}{X}{X}

Another user raised a concern regarding apps that cater primarily to one gender, leading to an expressed desire for these apps to incorporate additional considerations for other genders.
\qas{This is such a useful app and it has everything, literally... The only thing I would say could be improved is that the app is based around women and mainly women and maybe could have more \textbf{advertising for men.}}{SHEIN}{Play Store}

{\noindentpara{Socio-economic Status}{Finally, one major concern for users is the limitation caused by a user's socio-economic status and payment preference.
A number of users complained about the single biggest issue limiting their use of an app, which was a result of the app being too expensive.
We found, in this scenario, that economic status impacts a user's access to the app. 
When a user is unable or not willing to pay for premium subscription services, the user experiences a sense of exclusion despite their enjoyment of the app.}}

One user explicitly stated that premium features offered in the application are too costly for them.
\qas{Premium features are too expensive. Video/audio calls are very poor when one end has a slow device or poor connection.}{Telegram}{Play Store}
Moreover, another user explained that despite their love for a video streaming application, the fees for the app are far to expensive for the user to maintain.
\qas{This is a very good place to watch anything you want!! But the price is pretty expensive.....I would really love to keep watching my precious anime on here but I can't keep up with the payments}{Hulu}{Play Store}

\subsection{Accessibility}
Our data suggests two main subcategories  \emph{\textbf{visual}} and \textbf{\emph{audio}} that relate to accessibility inclusiveness in user feedback.
In particular, this category covers any user feedback regarding concerns about software accessibility (R: 23.2\%, A: 14.5\%, X: 14.4\%).

{\noindentpara{Visual}{Visual accessibility emerged as a significant subcategory with multiple users expressing challenges due to design choices that do not accommodate their visual needs.}}

One user on X mentioned how the teleconferencing application shared their sentiment of feeling neglect and lack of inclusiveness. The user feels that their accessibility concerns are unaddressed and pointed towards the psychological effects of inaccessible design.  
\qts{Feedback has been given to 
@Zoom about this innumerable times. At this point, the only thing I can take from this is that they don't care about how damaging their shitty captions are to \textbf{disabled and neurodivergent people.} All @Zoom does is enable microagressions.}{Zoom}{X}

Another user on Reddit described the eye strain and discomfort due to high contrast and overly dark mode themes. Users with eye conditions such as astigmatism described how the new design physically harms their experience. 
\qrs{I don’t get why ruining the oerfectly balamced dark mode by making it too dark was on top of discord’s to do list, but whoever came up with that idea should be fired asap for theur blunder. i can nolonger use the app for more than a minute before my eyes start to physically hurt, causing eyestrain. This could be considered ableist as \textbf{people with eye issues} are likely to be impacted in several ways, saw someone mention astigmatism for example.
my proposal is to let people costumise the color of the text and background of the UI, or at least give us the option to use the old darkmode Colors.}{Discord}{Reddit}
The issue with the messaging app was echoed by another user who complained that the app developers caused a huge accessibility issue with their new update. 
\qrs{A few months ago my app updated and I immediately realized that if there was no way to revert back to the old format, it would be impossible for me to keep using the app — I physically couldn’t look at it without it giving me horrible eye strain, the contrast was just WAY too high... It’s impossible for me to use the app now because it will destroy my eyesight; it \textbf{physcially hurts for me} to try to look at the app because of the \textbf{astigmatism.}
I’ve noticed Discord becoming more and more out of touch with its user base and this is really one of the biggest things that shows me that they’ve lost all care about the people they serve. I’m not trying to sound melodramatic, but they’ve quite literally caused a huge accessibility issue with their app.}{Discord}{Reddit}

Finally, one user review from the Play Store exemplified the issues they had with accessibility and registering for an account. 
The anti-spam system that the organization used failed to consider that a user may not be able to solve the visual challenges. 
\qas{So I put in my email user name password and Birthday, and it wanted me to confirm that I'm not a freaking robot! Well \textbf{I'm blind, so I couldn't do the visual challenge,} the text challenges are so hard I don't think anyone BUT a robot could figure them out, and it kept timing me out and taking me back. Their accessibility option wasn't working either. I'm dumping this app like the trash it is.}{Discord}{Play Store}

In all of these examples, users reported being excluded from the apps due to the lack of consideration for their visual needs. 
To offset these issues, app developers would need to consider designing more accessible user interfaces that consider these user needs.

{\noindentpara{Audio}{Similarly, numerous users complain about audio-related inclusiveness.}}

One user on X criticized the lack of support for deaf people in the messaging application. 
The application hid the voice-to-text bot behind a paywall upon a recent update.
\qts{As a \textbf{deaf person} voice messages are difficult, and they're a very used tool. So I had to adapt using voice to text bot, which suddenly stopped working after Telegram announced their update. Now it's behind a paywall.
It must be nice to get money restricting \textbf{disabled people} :)}{Telegram}{X}

Similarly, one user complained about how a video sharing application frequently hides subtitles after showing advertisements. 
In some situations, the option to even turn on subtitles disappears entirely, excluding those who are deaf to transcribing the audio. 
\qrs{I've noticed that YouTube loves to turn off subtitles for the videos, after getting back from an add. It's not always, but it happens A LOT!! It's ok to have ads, as I'm not paying to watch stuff, but for \textbf{deaf people like myself,} it can be pretty annoying, as after some ads, the option to turn on the subtitles again just disappears! I think should be addressed, cause why just the deaf needs to pay for the premium service, just to understand the videos we are watching whitout all this inconvenience? Anyone else have experienced that? What do you guys think about this subject?}{Youtube}{Reddit}

Additionally, one user reflected on the lack of visual captions for a video streaming application, which severely hurts users who have audio impairment and require captions to watch the videos. 
\qas{Wow. I hope you are not \textbf{DEAF.} Because there Closed Caption for their programs is borderline criminal. I should file a complaint with the ADA. The smallest font possible, using times new Roman characters which are more difficult to read(why do you think road signs use a font similar to calibri). And all the text is matte black. It's like they said "let's take everything good about other streaming services, and use none of it." Paramount plus, the spirit airlines of streaming apps.}{Paramount +}{Play Store}

\subsection{Other Human Values}
The final category in our taxonomy relates to user feedback about violations of different basic human values.
This category was the least common out of the taxonomy (R: 10.4\%, A: 13.3\%, X: 17.1\%). 
Basic human values refer to ``principles that guide social life and are modes of conduct that a person likes or chooses among different situations'' \cite{parashar2004perception}.
To better structure and present such related user feedback, we draw upon Shalom H. Schwartz's \cite{schwartz1992universals} theory of basic human values.
The theory amalgamates 58 values grouped into 10 categories. 
In our dataset, we identified user concerns related to 3 values \textbf{\emph{freedom}}, \textbf{\emph{social justice}}, and \textbf{\emph{benevolence}}.
We identified users expressing exclusion from the app when they perceive a violation of these human values.

{\noindentpara{Freedom}{
This sub-category particularly advocates for ``freedom of thought or speech,'' often tied to ideals of equality.
We found users describing their feelings of exclusion from freely providing their thoughts or interjecting their voices.
Many users express frustration about being restricted from voicing their opinion in any given app.}}

For example, one user on X complained that a social media app by default prevents marginalized voices from accumulating viewers. 
\qts{Apparently, Instagram has set everyone’s account to default ``limit sensitive content”, which is a large reason so many \textbf{marginalized voices} aren’t receiving engagement or being seen. 
}{Instagram}{X}

Similar frustrations were echoed by users who felt that some social media apps were censoring based on their speech. 
\qas{I remember when Facebook was fun, not it's not any longer. It was a place to connect with family and friends. A place to talk and share thoughts and ideas but since it started to \textbf{censor speech} it's no longer a place to be.}{Facebook}{Play Store}
All these examples suggest that users with specific opinions may feel marginalized or excluded from these platforms as their ability to freely express themselves is being restricted.

{\noindentpara{Social Justice}{
We also found user exclamations regarding social justice, leading users to experience a sense of exclusion within the app.
Social justice refers to a commitment to ensure that all members of society, irrespective of their race, religion, gender, or other characteristics, have equal access to opportunities and resources \cite{schwartz1992universals}, in this context, equal access to the app and all the features.}}

For instance, one user on Reddit explained how they reported other users for violating terms of services for making highly inappropriate comments, yet, those accounts do not face any restrictions. 
\qts{I keep reporting really vile things, and they keep saying it's not a violation of their TOS, but the few times I've come back at those people with something like the Caucasian refuse insult, immediate 3 day time out for violating the TOS on hate speech. Why aren't they more specific that said rules protect one race ad none of the others?}{Facebook}{Reddit}

In another example, a user of a mobile payment application cites a lack of equal treatment based on their birth name, leading to an inability to access the app's features.
\qas{This company is blatantly Racist! I opened a new account for the first time so I could collect a giftcard. They automacticly made it a restricted account because my \textbf{birth name is Arabic! I am black and American by birth!} I made repeated attempts through their customer support to fix the problem, but they refused to unrestrict my account. So now I can't collect my giftcard! DON'T USE VENMO!}{Venmo}{Play Store}

{\noindentpara{Benevolence}{
This sub-category pertains to users feeling excluded due to their concern about the well-being of people they interact with regularly, i.e., family and friends \cite{schwartz1992universals}
We particularly found instances where users believe the software lacks child-friendly content and lacks features that could improve the overall family experience.
}}

For example, one Reddit user detailed their concern for their kids and their use of a popular video sharing application. 
Their concern stems from the unrestricted number of disturbing content on the platform that is easily accessible by young children.
\qrs{\textbf{My young kids} love watching YouTube. But I'm really uncomfortable letting them have unfettered access to the full YouTube content. There is a lot of disturbing content and the algorithms are known for luring people into extreme content.  I have Roku TVs and the YouTube for Kids app is unavailable. ...}{Youtube}{Reddit}

Another user from the Play Store complained about their perceived worry about inappropriate content for young children on a leading video streaming application. 
\qas{... The recent video released of Disney internal meets proves they dont have the best interest of \textbf{children} in mind. When you decide to make wholesome content again and stop pushing sexuality on children is when we will spend money with you again. From now on our home will be Disney free. ...}{Disney+}{Play Store}

\begin{mybox}
RQ1: We find five major categories of inclusiveness, ranked in order of prevalence: \emph{algorithmic bias}, \emph{technology}, \emph{demography}, \emph{accessibility}, \emph{other human values}, and which we present in the form of a taxonomy of inclusiveness. 
\end{mybox}

\begin{table*}[h!]

    \centering
    \caption{Total number of inclusiveness-related user feedback in the 5 types of Apps from Reddit, Google Play Store, and X}
    \label{table-stat-category}
    \begin{tabular}{p{1.3cm}cccccc} \toprule
    Source & Technology  & Algorithmic Bias	& Demography	& Other Human Values	& Accessibility & Total \\\midrule
   Reddit & 124 	& 102	& 92 & 50	&	111 &  479 \\  
     Play Store &  67	&	68 & 49 &	34 &	37 & 255   \\
     X &  23	& 15	& 20 &  18	&	 16 &  92	 \\ 
      Total & 214	& 185	&  161 & 102	&	164 &  826 \\  \midrule

    \end{tabular}
\end{table*}
\section{RQ2: Inclusiveness across different Sources of User Feedback}
To answer our second research question, \textit{``How does inclusiveness-related user feedback differ across different sources of feedback?"} we analyzed the distribution of the categories in our taxonomy across the three sources.
As shown in Table \ref{table-stat-category}, \emph{Reddit} has a far greater number of inclusiveness user feedback in comparison to the other two sources.
After all, \emph{Reddit} has by far the largest character limit among the sources and is a platform for users to engage in long form discussion.
Therefore, users often share a significant amount of details, including inclusiveness problems and the entire backstory to their exclusion. 
Users who post on Reddit can engage in through discussions with other online users about their inclusiveness issues.
Particularly, we found the inclusiveness feedback frequently occurring for \emph{technology} and \emph{accessibility} categories, where users are likely to give detailed descriptions of their poor support for their diverse needs such as audio or visual qualities.

In contrast, we observe slightly different popular categories for \emph{Play Store}. 
We discovered that \emph{algorithmic bias} is a more prominent category for \emph{Play Store}, which makes sense as users who encounter unfair treatment related to terms of service or biased AI recommendations in an app can directly voice their reviews to warn other potential app users. 
For \emph{X}, the popular categories differ slightly, and \emph{technology} is by far the common category.
Since \emph{X} is most often a platform for users to contact support or voice real time feedback, users will discuss device or network \emph{technology} problems when they occur.

\begin{mybox}
RQ2: Reddit contains more inclusiveness feedback in comparison to X and Google Play Store. Depending on the source of feedback, users express different kinds of categories of inclusiveness feedback.
\end{mybox}

\section{RQ3: Automated Identification of Inclusiveness user feedback} 
In answering our third research question, \textit{``How effective are large language models in automatically identifying inclusiveness-related user feedback?"} we assessed the effectiveness of using GPT4o-mini as the model was recently released by OpenAI and it supports fine-tuning, as detailed in Section \ref{sec: classification-method}.
We used three different approaches on the top of the model for classification including zero shot, few shot, and fine-tuning, and measured the performance in terms of precision, recall, and F1-score. We use the macro version of the metrics as the data is imbalanced~\cite{li2010data}. Since we focus on the class of \textit{Inclusiveness}, hence, we consider the Recall of the \textit{Inclusiveness} class as our main evaluation metric rather than the F1-score.

The results are outlined in Table \ref{table-classification-result}.
We found that the best overall macro average was for zero-shot, where the precision was 0.95, the recall was 0.80, and the F1-score was 0.86.
Among the three approaches, zero-shot achieved the highest precision (0.92) for inclusiveness, however, this was achieved at the cost of a lowered recall (0.61) for inclusiveness. 
This result suggests that zero-shot learning has limited sensitivity to inclusiveness-related feedback.
For few-shot learning, inclusiveness recalls improved significantly to 0.72, indicating enhanced sensitivity to inclusiveness-related feedback, but at the trade-off of precision (0.45). 
Finally, fine-tuning achieved the lowest macro average F1-score, however, it also achieved the highest inclusiveness-related recall at 0.78. 
This showed that the fine-tuned model can capture most inclusiveness-related feedback. 
The trade-off is that precision drops to 0.28, but we can identify more true positive inclusiveness feedback.
Since the dataset is heavily imbalanced towards non-inclusiveness feedback, the number of user feedback incorrectly labeled as inclusiveness by the model is low overall. 
Given that our purpose is to identify inclusiveness-related user feedback using large language models, our results indicate the fine-tuning approach achieves the best performance for this goal.

\begin{table}[th!]

    \centering
    \caption{Results of Classifying between Inclusiveness and Non-Inclusiveness using GPT4-o Mini}
    \label{table-classification-result}
    \begin{tabular}{p{2.0cm}cccc} \toprule
      Source &  Class  &  Precision & Recall & F1-Score  \\\midrule
           \multirow{2}{*}{\shortstack[l]{Zero shot}} &  Inclusiveness & 0.92 & 0.61 & 0.73       \\ 
           &   Non-Inclusiveness    &   0.99    &  0.99       & 0.99     
    \\  
    &   Macro Avg.    &   0.95    &  0.80       & 0.86     
    \\ \midrule
      
     \multirow{2}{*}{\shortstack[l]{Few shot}} &  Inclusiveness & 0.45 & 0.72 & 0.55       \\ 
           &    Non-Inclusiveness    &   0.99    &  0.98       & 0.99     
    \\  
    &   Macro Avg.    &   0.72    &  0.85       & 0.77     
    \\\midrule
 \multirow{3}{*}{\shortstack[c]{Fine tuned}} &  Inclusiveness & 0.28 & \textbf{0.78} & 0.41       \\ 
           &    Non-Inclusiveness    &   0.99    &  0.95       & 0.97     
    \\  
    &   Macro Avg.    &   0.64    &  0.86       & 0.69     
    \\\bottomrule
    \end{tabular}
\end{table}

\begin{mybox}
RQ3: Fine-tuning achieved the highest recall, which indicates it is the most proficient model for identifying inclusiveness-related user feedback, but at the trade-off of increased false positives. Whereas, zero shot and few shot learning have lower recall and are less proficient in identifying inclusiveness-related user feedback.
\end{mybox}

\section{Discussion}
Inclusiveness-related user feedback is often expressed by users on Reddit, Play Store, and X. 
Our study proposed a taxonomy of inclusiveness-related user feedback comprised of 5 categories algorithmic bias, technology, demography, accessibility, and other human values.
These categories demonstrate how end users are excluded from using an app or feature due to being part of different groups such as age, economic status, or disability.
Previous work by Khalajzadeh \emph{et al.} \cite{khalajzadeh2022supporting} provided a preliminary introduction to inclusiveness concerns in open source apps from GitHub and Google Play Store, but open-source software represents a small fraction of the total number of user facing apps.
In contrast, we explore a more diverse and comprehensive range of highly popular commercial apps, specifically focusing on the end users' perspective, finding a significantly greater number of inclusiveness-related user feedback than prior work.

\subsection{Importance of Inclusiveness for Requirements}
Traditionally, user feedback analysis focuses on finding bugs and feature requests, which may not always include inclusiveness-related issues \cite{maalej2015bug}. 
Yet, inclusiveness-related user feedback unveils the often overlooked requirements that are necessary to support broader audiences' ability to use an app.
Requirements and by extension software design may suffer from only focusing on an \emph{average} user \cite{szlavi2023gender}
\cite{agrawal2024development}, and ignoring the needs of diverse end users. 
For example, one of the key categories that emerged through our analysis was \textit{algorithmic bias}. 
With the rapid evolution of AI, companies are increasingly inclined towards integrating algorithms in their decision-making process.
Many features are automated by AI decision-making, but these AI systems are often flawed due to underrepresented data and biased methods \cite{akter2021algorithmic}, causing bias towards certain groups of users \cite{brun2018software}. 
For instance, on a popular social media platform, AI decides if content is allowed on their platform based on Community Standards \cite{facebook_community}. 
Our analysis revealed many users complained about being banned from the platform due to policy violations without warning or prior notification and perceived the platform is biased towards the group they belong to. 

\vspace{2mm}
\qrs{It appears facebook only care to remove posts reported by people form the 1st world and posts written in English even though there is a fairly accurate auto translate feature. \textbf{I live in europe, Macedonia} ... Furthermore, the post I made this topic about had a video in it, you don't need to know the language to understand and see what is going on.}{Facebook}{Reddit}
\vspace{2mm}

Other famous incidents exposing biased AI systems include the COMPAS recidivism algorithm \cite{larson_compas_algorithm}, which had a significantly greater likelihood of incorrectly judging black defendants in comparison to white defendants, whereby black defendants were more likely flagged as high risk. 
Recently, a social media platform agreed to a settlement after it was revealed it implemented features in its advertising to exclude specific groups of people \cite{tobin_facebook_2022}.


We also found that \textit{demography inclusiveness} was another major category. 
For example, in the \textit{age} subcategory, we found seniors struggling
to adapt to user interface changes and motor and non-motor abilities with different apps. 
Specifically, the ability to see the small fonts used on the apps was limited by their age and many apps did not support software design that catered to these specific needs. 
Moreover, some features in various applications were often hidden in obscure placements, wreaking confusion on the user experience for young children or elderly users. 
Our identified taxonomy of inclusiveness-related user feedback is a step towards alleviating these issues. 
It can help organizations transcend beyond just bugs or features and identify critical inclusiveness-related issues that can be aware and translated into requirements.

\subsection{Implications for Practitioners}
Previous research has indicated that developers are predominantly male, technically skilled, and affluent \cite{li2015makes}, therefore potentially having a substantial gap in understanding the diverse end-users they serve. Awareness of the inclusiveness issues that we identified in our study will enable them to learn and consider the diverse user needs and as a result, develop more inclusive software. Furthermore, studies have employed automated identification of bugs and features from app reviews \cite{maalej2015bug}. 

The automated approach proposed by our study provides potential solutions in the form of automated flagging (i.e., a plugin) on source platforms to address the limitations of the current manual approach in identifying inclusiveness-related user feedback.
As indicated in our taxonomy, algorithmic bias is a prominent issue of inclusiveness from the end user perspective. 
The study of reducing bias in machine learning systems has been a large subject area \cite{chakraborty2021bias}. 
Several studies explore reducing bias in AI systems, particularly for those that conduct automated decision-making \cite{krishnan2014methodology, tsintzou2018bias}. 
These studies attempt to research how to minimize algorithmic bias by improving the quality of the training dataset, model training, and testing.
However, our study indicates the value of leveraging inclusiveness-related user feedback to understand exclusions that users face when using software, which can mitigate and reduce unintentional exclusions caused by AI-assissted decision making systems. 

\subsection{Implications for Researchers}
Our findings carry several implications for future work. First, our results reveal persistent issues with algorithmic bias, particularly in recommendation systems and content moderation. These biases disproportionately impact some groups of people, highlighting the need for further research to mitigating bias in automated AI decision making systems. Second, our research emphasizes the need to explore the intersectional issues where inclusiveness categories such as age, socio-economic status, and location intersect to compound exclusion in an app or feature. 
While our study demonstrates the potential of large language models like GPT-4o Mini in automating inclusiveness feedback detection, the trade-offs between precision and recall highlight a gap for further exploration. We advocate future research in improving automated tools to ensure accurate detection of inclusiveness-related issues at scale.
Finally, the apps we analyzed have millions of users from different cultural backgrounds. 
According to Hofstede's Cultural Dimensions Theory \cite{hofstede1991cultures}, people generally belong to either  \emph{individualistic} or \emph{collectivistic} culture, which may influence their expectations and interactions with technology.
As technology often reflects the cultural characteristics of the society in which it is developed \cite{al2007learning}, users from different cultural backgrounds may expect different functionalities.
From our taxonomy categories like \textit{demography} and \textit{other human values}, we notice the potential impact of ethnic culture and user preferences. This can be a fruitful direction for future research as the cultural impact on system usage and end-user preferences is largely unexplored. 

\subsection{Threats to Validity}
We describe several threats and mitigation strategies in our study using the total quality framework of Roller \cite{roller_applied_2015}.

\emph{Credibility} indicates ``the completeness and accuracy associated with data gathering'' \cite{roller_applied_2015}. 
This study may have the threat of sampling bias because we collected user feedback from 50 apps from the sources of feedback.
However, we selected a diverse group of apps, and the feedback sources are also common platforms that users often use to discuss concerns.
Our study also used standard web scraping libraries to collect the data.
Additionally, we try limiting bias by creating a randomly sampled batch of user feedback to conduct manual annotation. 
We did not seek to give more weight to any particular app or source of feedback.

\emph{Analyzability} refers to ``completeness and accuracy related to the processing and verification of data'' \cite{roller_applied_2015}.
There is a potential limitation from the annotators misinterpreting implicit information from the data.
To mitigate the threat, we analyzed the data with two co-authors who followed a social-technical grounded theory approach \cite{hoda2021socio}, where open coding, constant comparison, and memoing were used to analyze the feedback for inclusiveness. 
The co-authors were in constant dialogue during the coding process to ensure consistency and remove bias.
Since this study leverages user feedback from three popular sources (i.e., Reddit, X, and Google Play) and different apps, we were able to triangulate our analysis with the different sources.
The constant communication also helped to ensure that they had a shared understanding of the inclusiveness taxonomy.
Additionally, we only used GPT4o-mini for inclusive feedback identification. However, GPT4o-mini is one of the most recent models released by OpenAI.

\emph{Transparency} refers to the ``completeness of
the final documents and the degree to which the research can be fully evaluated and its transferability'' \cite{roller_applied_2015}.
We provide extensive and rich descriptions of our methodology, as well as detailed quotes to support our taxonomy.
We also release the entirety of our data in our replication package, including our manually labeled dataset \cite{replication_package}. 


\emph{Usefulness} specifies the ``ability to do something of value with the research outcomes'' 
\cite{roller_applied_2015}.
Our study shed more insight into the role of inclusiveness in user feedback. 
More importantly, we advance the state of knowledge of inclusiveness by providing a taxonomy for the different types of inclusiveness-related discussions. 
In particular, our study encompasses a significant number of user feedback and includes more empirical insights for organizations.
We acknowledge that our results may not hold for every software app depending on its functionality, but we believe organizations can benefit from the inclusiveness categories as they try to consider the concerns of diverse end users.

\section{Conclusion}
Our study follows a socio-technical grounded theory approach to gain a deeper understanding of inclusiveness-related user feedback from end users. 
Across manual analysis of over 22K user feedback posts from Reddit, X, and Google Play Store regarding 50 popular for profit apps, we build a taxonomy of inclusiveness.
Our taxonomy has five main categories, including algorithmic bias, technology, demography, accessibility, and other human values. 
The large language models that we evaluated on our data show that we can automatically identify inclusiveness-related feedback among general user feedback. 
Our results indicate to practitioners that these online sources contain a rich source of inclusiveness feedback that organizations should consider to build more inclusive software products for diverse end users. 
We also present our labeled dataset that researchers can use to refine tooling to better support practitioners.


\bibliographystyle{ACM-Reference-Format}
\bibliography{main}

\appendix

\end{document}